\documentclass[journal]{IEEEtran}

\usepackage{amsmath}
\usepackage{amssymb}
\usepackage{graphicx}
\usepackage{booktabs}
\usepackage{multirow}
\usepackage{url}
\usepackage{cite}
\usepackage{xcolor}

\begin{document}

\title{Template Ageing and Longitudinal Verification in Fixed-Text Keystroke Dynamics: A Subject-Disjoint Study Across Eight Weeks}

\author{Simon~Parkinson,
        Saad~Khan,
        Na~Liu,
        Qing~Xu
\thanks{Prepared for review.
Simon Parkinson, Na Liu, and Saad Khan are with the Department of Computer Science,
University of Huddersfield, HD1 3DH Huddersfield, U.K. (e-mail: s.parkinson@hud.ac.uk),
Qing Xu is with College of Intelligence and Computing, Tianjin University, Tianjin, China}}

\markboth{IEEE Transactions on Biometrics, Behavior, and Identity Science}{}

\maketitle

\begin{abstract}
Behavioural biometric templates are widely believed to degrade as the gap between enrolment and verification grows. However, few studies measure this \emph{template ageing} effect directly under controlled conditions. In this study, we collected a longitudinal dataset of 40 fixed passwords, each typed four times per weekly session over eight consecutive weeks. We then compare a scaled-Manhattan matcher (M1), a gradient-boosted classifier (M2), a TypeNet-style recurrent embedding model (M3), and a TypeFormer-style Transformer (M4) under a 5-fold subject-disjoint protocol and a single test design that jointly varies the mechanism and the enrolment gap to the query $\Delta \in \{0,\ldots,7\}$ weeks. Template ageing proves large and systematic. The verification error increases monotonically with $\Delta$ for every mechanism, from an EER of 14.6--27.2\% at $\Delta{=}0$ to 25.5--37.1\% at $\Delta{=}7$, or 1.7 percentage points of decision error per week elapsed ($p<0.001$), with error at $\Delta{=}7$ exceeding that at $\Delta{=}0$ in every fold without exception. However, the choice of mechanism matters more than its rate of ageing. Baseline accuracy spans 12.6 percentage points across the four mechanisms, the degradation each accumulates over seven weeks spans only 2.3 points, and ageing never reorders them, the enrolment-time ranking holding at every gap. A matcher can therefore be chosen on same-session accuracy, with ageing managed by re-enrolment scheduling rather than by matcher selection. The two properties are nonetheless distinct, as M3 is the least accurate mechanism yet ages significantly more slowly than M1 under every specification tested. We also find that the influence of training randomness differs sharply by architecture, with 58\% of the recurrent model's fold-to-fold variance attributable to seed noise against 19\% for the Transformer. Because the smaller ageing-rate differences are correspondingly sensitive to modelling choices, while the accuracy differences and the ageing effect are not, we recommend that comparative ageing-rate claims be supported by seed-level score fusion, independent replication, and an alternative outcome-model specification.
\end{abstract}


\section{Introduction}
\label{sec:intro}

Password-based authentication remains the dominant access-control mechanism, despite well-documented tensions between the security benefits of complex password policies and the usability costs they impose on users~\cite{habib2018user,shay2016designing}. Keystroke dynamics is the passive capture of key-press and key-release timing as a user types and has long been proposed as a low-cost behavioural biometric that can be added to an existing password system without changing user behaviour or requiring additional hardware\cite{teh2013survey}.

Where keystroke dynamics is deployed this way, a template is typically enrolled once and then used for a period of time without re-enrolment, which could be weeks or months. If verification accuracy degrades over that period, and does so at different rates depending on the matching mechanism, the choice of mechanism has consequences for a deployment that a same-session accuracy comparison alone cannot reveal. A system that performs best on day one is not necessarily the one that is the most reliable over a prolonged period. Whether such comparative claims can even be trusted from a single experimental run is itself an open methodological question, and one that this paper treats as being as important as the substantive comparison.

There are four gaps that motivate this study. First, template ageing is widely acknowledged and several adaptive mechanisms already exist to counteract it~\cite{kang2007continual,mhenni2019double,giot2011analysis,pisani2016enhanced,yang2021detection}. However, this body of work is orientated towards managing ageing rather than isolating the ageing effect itself. Second, population-level embedding models trained with a metric-learning objective, including TypeNet~\cite{acien2022typenet} and its Transformer successor TypeFormer~\cite{stragapede2024typeformer}, are now the leading keystroke techniques. However, they have so far only been demonstrated on free-text data rather than longitudinal fixed-text passwords, and have never been evaluated to see how their accuracy changes as the enrolment-to-query gap grows. Third, longitudinal keystroke studies typically report raw percentages, even though repeated observations nested within participants and passwords are a natural fit for mixed-effects modelling~\cite{baayen2008mixed}. This kind of modelling is rarely applied in this domain. Fourth, comparative claims about which mechanism ages fastest are typically reported from a single trained model and a single statistical specification. There is usually no check on whether the result would survive retraining, or whether it would still hold under an equally defensible alternative choice of outcome model.

We address these gaps using a dataset of 85 participants who each typed 40 passwords of varying lengths and substitution types, with four repetitions per password within a single weekly session, over eight consecutive weeks~\cite{parkinson2023empirical}, within a subject-disjoint protocol (training and testing on non-overlapping participants) that treats the enrolment-to-query gap as a controlled variable.

The paper's contributions are as follows: 
\begin{itemize}
    \item a measurement of how much fixed-text password verification degrades as the time between enrolment and use grows, tracked week by week across eight weeks;
    \item a single trial design under which four matching mechanisms, including a distance-threshold matcher, a gradient-boosted classifier, and recurrent and Transformer embedding models, are compared so that differences in accuracy, differences in ageing rate, and any interaction between the two are all from the same set of trials;
    \item an analysis that accounts for repeated trials coming from the same participants and the same passwords, providing effect sizes and confidence intervals;
    \item an estimate of how much of the apparent difference between mechanisms is simply training-run randomness, obtained by retraining each model several times on identical data;
    \item a demonstration that comparative-ageing-rate claims can also depend on the choice of statistical outcome model itself, not only on training randomness, by checking whether each finding survives an equally defensible alternative specification.
\end{itemize}

The remainder of the paper is structured as follows. Section~\ref{sec:related} reviews related work. Section~\ref{sec:methodology} details the methodology of this study. In Section~\ref{sec:results}, the results are reported. Section~\ref{sec:discussion} provides a discussion, focusing on what the evidence supports. Finally, in Section~\ref{sec:conclusion} a conclusion is provided.

\section{Related Work}
\label{sec:related}

\subsection{Keystroke Dynamics Fundamentals}
\label{sec:fundamentals}

Keystroke dynamics systems are categorised as fixed-text or free-text~\cite{teh2013survey,shadman2023keystroke}. Fixed-text systems typically derive dwell, flight/digraph, and trigraph timing features from key-press/release timestamps~\cite{banerjee2012biometric}. It is widely acknowledged that no single feature set is universally optimal across passwords or users~\cite{balagani2011discriminability}. Sae-Bae and Memon~\cite{saebae2022distinguishability} propose a metric predicting a template's false-acceptance contribution from user-specific feature variation without impostor data. This is complementary to our participant-level random effect (Section~\ref{sec:statmodel}), which models this same variability rather than predicting a per-template score.

\subsection{Matching Mechanisms}
\label{sec:mechanisms}

Distance-threshold matchers remain popular for transparency and remain competitive with more complex anomaly-detection algorithms on small datasets~\cite{killourhy2009comparing}. Previous work using classifier ensembles reaches sub-1\% EER on fixed-text data when competence is modelled explicitly~\cite{porwik2021dynamic}. The state of the art is built around learnt embedding models with metric-learning objectives. More specifically, Ayotte et al.~\cite{ayotte2020fast} demonstrated an instance-based deep approach to free-text verification. In other work, TypeNet~\cite{acien2022typenet} trains a recurrent encoder with a triplet loss over a population of typists, enrolling identities from a handful of samples via gallery-distance comparison. In more recent work, TypeFormer~\cite{stragapede2024typeformer} has replaced the recurrent encoder with a Transformer using Gaussian range encoding, further reducing the equal error rate (EER) with as few as five enrolment sessions. Deep learning carries documented risks here, such as adversarial vulnerability~\cite{papernot2016limitations} and reduced interpretability~\cite{samek2017explainable}. These motivate a transparent baseline for comparison. The architecture/objective landscape continues to broaden beyond the TypeNet/TypeFormer pair and triplet loss evaluated. Momeni and BabaAli~\cite{momeni2023freetext} compare Transformer variants and loss functions for free-text authentication. Angular-margin losses (ArcFace~\cite{deng2019arcface}) and supervised contrastive objectives~\cite{khosla2020supervised} are widely-used alternatives to triplet loss. Self-supervised pre-training has also been demonstrated for gait~\cite{pincic2022gait} and mouse dynamics~\cite{zhang2024mouse2vec}, although not, to our knowledge, for fixed-text keystroke verification. We evaluate one recurrent and one attention-based model under the same triplet objective, prioritising a tractable, thoroughly-replicated comparison (Section~\ref{sec:ageingrates}). No published work, to our knowledge, applies the TypeNet/TypeFormer paradigm to a fixed-text, multi-week, multi-password dataset, or examines how such a model's accuracy changes as the enrolment-to-query gap grows.

\subsection{Longitudinal Effects and Password Characteristics}
\label{sec:longitudinal}

Keystroke accuracy under repeated use of the same phrase generally stabilises after a few repetitions, and longer passwords do not uniformly improve error rates~\cite{montalvao2015contributions,morales2016keystroke}. Both length and character substitutions also affect usable-security outcomes~\cite{bhana2020passphrase}. Studies combining length, substitution, and repetition within a pool over an extended multi-week period remain rare~\cite{parkinson2023empirical}. Previous work has examined repetition and template generalisability within individual weeks~\cite{parkinson2021password,parkinson2023repetition}. Widely used benchmark datasets, such as the GREYC keystroke~\cite{giot2009greyc}, typically involve few collection sessions rather than a sustained multi-week design.

\subsection{Template Ageing and Adaptive Biometric Systems}
\label{sec:templateageing}

A separate strand of research actively manages template degradation rather than measuring it. This includes continuous retraining~\cite{kang2007continual} and the double serial adaptation of both template and threshold~\cite{mhenni2019double}. Giot et al.~\cite{giot2011analysis} apply a semi-supervised update with multi-session evaluation, an approach that is close in spirit to our ageing curve but that measures how adaptation \emph{counteracts} degradation rather than isolating it. Pisani et al.~\cite{pisani2016enhanced} propose an Enhanced Template Update that uses impostor as well as genuine samples, and Yang et al.~\cite{yang2021detection} use a concept-drift framing that detects when drift has occurred rather than measuring its magnitude against elapsed time or mechanism. We draw a consistent distinction between two kinds of question here. Template-update work asks how to remain accurate once ageing is assumed, whereas raw ageing \emph{measurement} asks how large the unmitigated effect actually is and whether it depends on the mechanism used. This second question is a prerequisite for the first, as every number reported by an update evaluation already reflects the mitigation that was applied. It is also the gap that the protocol in Section~\ref{sec:trialdesign} is designed to address and against which future adaptive schemes could be benchmarked. There is a comparable literature for other modalities as well, including mouse dynamics~\cite{khan2024mouse}, which again prioritises adaptation over raw measurement.

``Ageing'' carries different meanings in biometrics. Physiological-permanence studies track degradation over months to decades. In one study, Yoon and Jain~\cite{yoon2015longitudinal} analyse fingerprint scores over up to twelve years across 15{,}000+ subjects, finding minimal degradation with controlled image quality. Our eight-week window measures the short-term behavioural drift (practice effects, motor variation, incidental environmental change such as Section~\ref{sec:limitations}'s hardware-transition covariate) rather than long-term physiological change. We retain ``template ageing'' as the established keystroke-literature term, but this result should be read as evidence that measurable drift begins early and grows monotonically from the first weeks, not as a claim about years of continued use.

\subsection{Statistical Methodology in Biometric Performance Evaluation}
\label{sec:statmethod}

Standard reporting (FMR, FNMR, EER) is mandated by ISO/IEC~19795-1~\cite{iso19795}, which specifies \textit{what} to report but not how to analyse repeated-measures data of this kind. Linear mixed-effects models fill that gap by providing proper standard errors, and are already established in repeated-measures behavioural research~\cite{baayen2008mixed}, though they are rarely applied to longitudinal keystroke evaluation. Bolle et al.~\cite{bolle2004error} show that naive FMR/FNMR confidence intervals overstate precision when scores from the same subject are correlated, and propose a subsets-bootstrap correction for this. Our participant-level bootstrap (Section~\ref{sec:metrics}) and our participant/password random effects (Section~\ref{sec:statmodel}) are two complementary responses to this same non-independence problem.

Table~\ref{tab:positioning} provides a summary how the design of the present study relates to this body of work.

\begin{table*}[t]
\centering
\caption{Positioning relative to selected prior work (qualitative; \checkmark{} = present, $\times$ = absent, $\sim$ = partially addressed).}
\label{tab:positioning}
\begin{tabular}{lccp{1.5cm}p{1.5cm}p{1.5cm}}
\toprule
Study & Longitudinal (weeks+) & Cross-week template test & Learned embedding model & Subject-disjoint evaluation & Inferential statistics \\
\midrule
Montalv\~{a}o et al.~\cite{montalvao2015contributions} & $\times$ & $\times$ & $\times$ & $\times$ & $\times$ \\
Acien et al. (TypeNet)~\cite{acien2022typenet} & $\times$ (free-text) & $\times$ & \checkmark{} & \checkmark{} & $\times$ \\
Parkinson et al.~\cite{parkinson2023empirical} & \checkmark{} & $\times$ & $\times$ & $\times$ & $\times$ \\
Giot/Pisani (template update)~\cite{giot2011analysis,pisani2016enhanced} & $\sim$ (multi-session) & $\sim$ (adaptation active) & $\times$ & $\times$ & $\times$ \\
Yang et al.~\cite{yang2021detection} & \checkmark{} (drift-detection) & $\times$ & $\times$ & $\times$ & $\times$ \\
This paper & \checkmark{} & \checkmark{} & \checkmark{} & \checkmark{} & \checkmark{} \\
\bottomrule
\end{tabular}
\end{table*}

\section{Methodology}
\label{sec:methodology}

\subsection{Dataset and Input Representation}
\label{sec:dataset}

This study is based on the dataset described in~\cite{parkinson2023empirical}. It includes 85 participants who typed 40 passwords drawn from English dictionary words at four nominal lengths (6, 8, 10, 12 characters), with five substitution categories (none, uppercase, numeric, symbol, combination), four times per session, once per week for eight weeks. Informed consent was obtained from each participant before acquisition.

The data was cleansed to remove incomplete submissions and submissions where unexpected events took place, such as a user moving the cursor back to edit what they had typed. This left 94{,}894 clean samples from 99{,}607 raw attempts. Missing participant-weeks are not imputed. The trial protocol (Section~\ref{sec:trialdesign}) does not have a trial for affected weeks, so participants with incomplete coverage contribute proportionally fewer trials (Section~\ref{sec:metrics}).

Two input representations are used. The first, a \textbf{raw sequence representation} used by the embedding models (Section~\ref{sec:embedding}), is the chronologically-ordered sequence of $2n$ key-event timestamps for an $n$-character password, with each timestamp expressed as an inter-event duration plus a press/release indicator. The second, an aggregated feature representation used by the classical and shallow-ML baselines (Section~\ref{sec:baselines}), concatenates full timing, dwell, press-to-press, release-to-press, release-to-release, and trigraph timings into a single $6n{-}6$-dimensional vector. The timing values are clipped on a 1.5$\times$IQR boundary, which is calculated only on the training-subject partition (Section~\ref{sec:protocol}) and then applied unchanged to the validation and test partitions. As an additional check, results are also reported without clipping, to test how robust each model is to raw timing noise.

\subsection{Subject-Disjoint Experimental Protocol}
\label{sec:protocol}

Embedding models (Section~\ref{sec:embedding}) are evaluated on participants unseen during training, to test generalisation. We use 5-fold subject-level cross-validation, in which participants are split into five folds of around 17 each. In each run, three folds (around 51 participants) are used for training, one fold is used for validation, which handles early stopping and threshold calibration, and one fold is held out for testing. All 40 passwords appear in every split. The metrics are aggregated across the folds, and the variance between the folds is explicitly reported rather than averaged away.

\subsection{Embedding Model Architecture}
\label{sec:embedding}

The primary model follows TypeNet~\cite{acien2022typenet}, using a bidirectional recurrent encoder that maps a raw event sequence to a fixed-length embedding and is trained with a triplet loss. Triplets pair an anchor and a positive sample drawn from two sessions of the same participant typing the same password, together with a semi-hard negative selected from a candidate pool of impostors. The semi-hard negative is the closest impostor sample that is still further from the anchor than the positive sample is. A second, higher-capacity model uses a Transformer encoder with Gaussian range encoding of inter-key timings, following TypeFormer~\cite{stragapede2024typeformer}, and is trained under the same triplet objective.

An enrolment gallery is the centroid of the embeddings from $k$ enrolment sessions, where $k \in \{1,3,5\}$ and $k{=}5$ is the primary configuration. The verification score is then the negative Euclidean distance from a query embedding to that centroid.

Table~\ref{tab:hparams} lists the hyperparameters shared by M3 and M4. These were fixed a priori from values reported in the TypeNet/TypeFormer literature and from brief manual trials on a single fold. As none of these values was tuned to any reported metric, the accuracy of both M3 and M4 (Section~\ref{sec:ageingresults}) should be read as a lower bound on what these architectures can achieve, not a ceiling (Section~\ref{sec:limitations}).

\begin{table}[t]
\centering
\caption{Architecture and optimisation hyperparameters for M3 (LSTM) and M4 (Transformer), fixed a priori and shared across folds and replicates.}
\label{tab:hparams}
\begin{tabular}{p{3.5cm}c}
\toprule
Hyperparameter & Value \\
\midrule
Hidden size & 64 \\
Embedding dimension & 64 \\
Encoder layers & 1 \\
Gaussian range-encoding bins ($n_\text{gaussians}$, M4 only) & 24 \\
Max duration for range encoding & 3.0 s \\
Triplet margin & 0.3 \\
Negative candidate pool size & 6 (semi-hard selection) \\
Optimiser & Adam, lr $=10^{-3}$ \\
Gradient clipping (norm) & 5.0 \\
Triplets per training step & 64 \\
Steps per epoch & 150 \\
Max epochs & 40 \\
Early-stopping patience & 6 epochs (val.\ loss) \\
Score-fusion seeds per fold (M3, M4) & 3 \\
\bottomrule
\end{tabular}
\end{table}

Motivated by the training-seed instability quantified in Section~\ref{sec:seedcheck}, both M3 and M4 are trained three times per fold with different seeds, and each reported score is the mean of the three. Embeddings are not averaged directly. This is because independently-trained triplet-loss networks have no constraint tying their coordinate systems together, so averaging raw embeddings could cancel genuine signal, whereas averaging the final distance scores each model produces for the same (gallery, query) pair is a standard, well-founded fusion technique.

\subsection{Unified Verification Protocol}
\label{sec:trialdesign}

The mechanism comparison and the ageing question are evaluated within a single trial design. For a test participant $p$, password $w$, enrolment week $i$, and query week $j$ with $i \le j$, a genuine trial compares the gallery built from the most recent $k$ sessions up to week $i$ against $p$'s own query sample from week $j$. An impostor trial compares that same gallery with every other test participant's session at week $j$. Because a fold contains around 17 participants, each genuine trial can be compared with up to 16 impostor trials, giving an impostor-to-genuine ratio that ranges from around 3.5 to 1 at $\Delta{=}7$ to around 14.8 to 1 at $\Delta{=}0$. We do not rebalance this ratio, since EER and FNMR are both derived from an ROC-style threshold search (Section~\ref{sec:metrics}) rather than from a statistic that is sensitive to class balance. The temporal gap $\Delta = j-i \in \{0,\ldots,7\}$ is recorded for every trial. Whenever there are two or more sessions for an enrolment week, one session of that week is always reserved outside the gallery, so that $\Delta{=}0$ trials are not deprived of genuine query material. This produces, for every mechanism, an EER surface that is jointly indexed by $\Delta$ and the mechanism, from which the effects of ageing, the effect of the mechanism and their interaction can all be estimated from the same trial set. The trial grid forms an upper triangle on $(i,j)$ pairs with $j \ge i$. Each diagonal of this grid groups trials from many different pairs of enrolment/query weeks that share the same $\Delta$, so the $\Delta{=}0$ diagonal, which spans 8 cells, contains many more trials than the single $\Delta{=}7$ corner cell (376{,}387 trials versus 51{,}729, Table~\ref{tab:eer}).

\subsection{Baselines for Comparison}
\label{sec:baselines}

Four matching mechanisms are evaluated under the identical trial design of Section~\ref{sec:trialdesign}:

\begin{itemize}
\item \textbf{M1, the classical baseline:} a \emph{scaled} Manhattan distance-threshold matcher on the aggregated feature representation (Section~\ref{sec:dataset}). The absolute difference of each feature from the centroid of the gallery is divided by its variability, taken from the gallery's own sessions, or from the training population if the gallery is too small to estimate the variability reliably, before summing, following the approach of Killourhy and Maxion~\cite{killourhy2009comparing}.
\item \textbf{M2, the shallow-ML baseline:} a gradient-boosted classifier trained on the absolute pairwise differences between two feature vectors, predicting whether a pair is genuine or an impostor, following a similar approach to~\cite{porwik2021dynamic}. A classifier is trained per password, unlike M3 and M4, which share a single representation across all 40 passwords. This is a deliberate choice that makes M2's task easier, and we flag it as a limitation on this specific comparison in Section~\ref{sec:limitations}.
\item \textbf{M3, the recurrent embedding model:} the TypeNet-style bidirectional LSTM described in Section~\ref{sec:embedding}.
\item \textbf{M4, the Transformer embedding model:} the TypeFormer-style encoder described in Section~\ref{sec:embedding}.
\end{itemize}

\subsection{Evaluation Metrics and Reporting Standards}
\label{sec:metrics}

For every (mechanism, $\Delta$) combination, we report EER and FNMR at fixed FMR operating points of 1\% and 0.1\%, following ISO/IEC~19795-1~\cite{iso19795}. Each EER pools every trial across test participants and passwords for that cell and runs a single ROC-style threshold search, rather than averaging separate per-participant EERs. This means that a participant or password that contributes more trials has proportionally more influence on the result, but this pooled convention matches ISO/IEC~19795-1 and minimises sampling variance, so we use it throughout. We checked whether this choice actually matters by running a per-participant-balanced re-analysis at $k{=}5$, clipped. It shifts every cell by at most 1.7 points and never changes the ordering of the mechanisms (see the supplementary material), so we do not believe it affects any conclusion drawn in this paper. The confidence intervals use bootstrap sampling at the participant-level within each fold, taking the 2.5 and 97.5 percentiles, which respects the fact that trials that share a participant are not independent~\cite{bolle2004error}.

\subsection{Statistical Modelling}
\label{sec:statmodel}

The results per-trial are modelled with a linear mixed-effects model~\cite{baayen2008mixed} rather than with aggregate percentages alone. Since EER is an aggregate quantity rather than a per-observation one, we instead model each trial's decision outcome at a calibrated threshold. This threshold is calibrated separately for each combination of fold, mechanism, and $\Delta$, using the EER point of the validation participants of that fold in the same $\Delta$, and then applied to the corresponding test trial to produce a binary outcome (\texttt{is\_error} $\in \{0,1\}$). Calibrating separately for each $\Delta$, rather than grouping across gaps, avoids conflating a miscalibrated operating point with the ageing effect itself, since the score distributions shift as $\Delta$ grows (Fig.~\ref{fig:ageing}). We use a \emph{linear} specification as our primary model for interpretability, so that its coefficients sit directly on the probability-of-error scale used throughout the paper. However, this choice is not safe by assumption. As shown in Section~\ref{sec:mixedeffects}, a logistic sensitivity check changes two of the three conclusions of the interaction of the mechanisms. We flag this prominently, as it affects how the central result of Section~\ref{sec:mixedeffects} should be read (Sections~\ref{sec:ageingrates} and~\ref{sec:limitations}). The specification is as follows.
\begin{align}
\texttt{is\_error} \sim{}& \Delta_c \times \text{Mechanism} + \text{Length}_c \nonumber \\
&+ \text{Substitution} + \text{HardwareChange} \nonumber \\
&+ (1 \mid \text{Participant}) + (1 \mid \text{Password}) \nonumber \\
&+ \text{Fold}
\end{align}
where $\Delta_c$ and $\text{Length}_c$ are mean-centred. The participant and password are entered as crossed random intercepts. The fold is treated as a \emph{fixed} effect rather than as a third crossed random effect because it is a fixed partition of participants induced by the cross-validation design rather than a sample from a population of folds. Treating the fold as random made the model numerically unstable in a repeated sample test, where the nested-model likelihood-ratio test swung from $p{=}1.0$ to $p{<}10^{-38}$ across subsamples of identical data. The interaction between $\Delta$ and the mechanism, which tests whether M3 and M4 degrade at a different rate than M1, is the central term in the model. Fixed effects are reported with values of $z$ and $p$, and nested models are compared using the likelihood-ratio test.

\subsection{Training-Seed Robustness Check}
\label{sec:seedcheck}

Training M3 and M4 only once per fold would combine genuine population variation with training-run variance arising from triplet sampling and weight initialisation. This is a material problem identified during development (Section~\ref{sec:ageingrates}), and motivates the score-fusion protocol described in Section~\ref{sec:embedding}. We directly quantify this residual variance from the training-run. Independently of fusion, the M3 and M4 of each fold are retrained three times with different seeds on identical data, and the pooled EER is recomputed without fusion each time. The within-fold, across-seed standard deviation is then compared against the across-fold standard deviation in the main results to estimate what fraction of the reported mechanism differences could be training noise and how much residual uncertainty three-seed fusion should be expected to leave behind.

\section{Results}
\label{sec:results}



\subsection{Accuracy and Template Ageing}
\label{sec:ageingresults}

\begin{figure*}[t]
\centering
\includegraphics[width=0.65\textwidth]{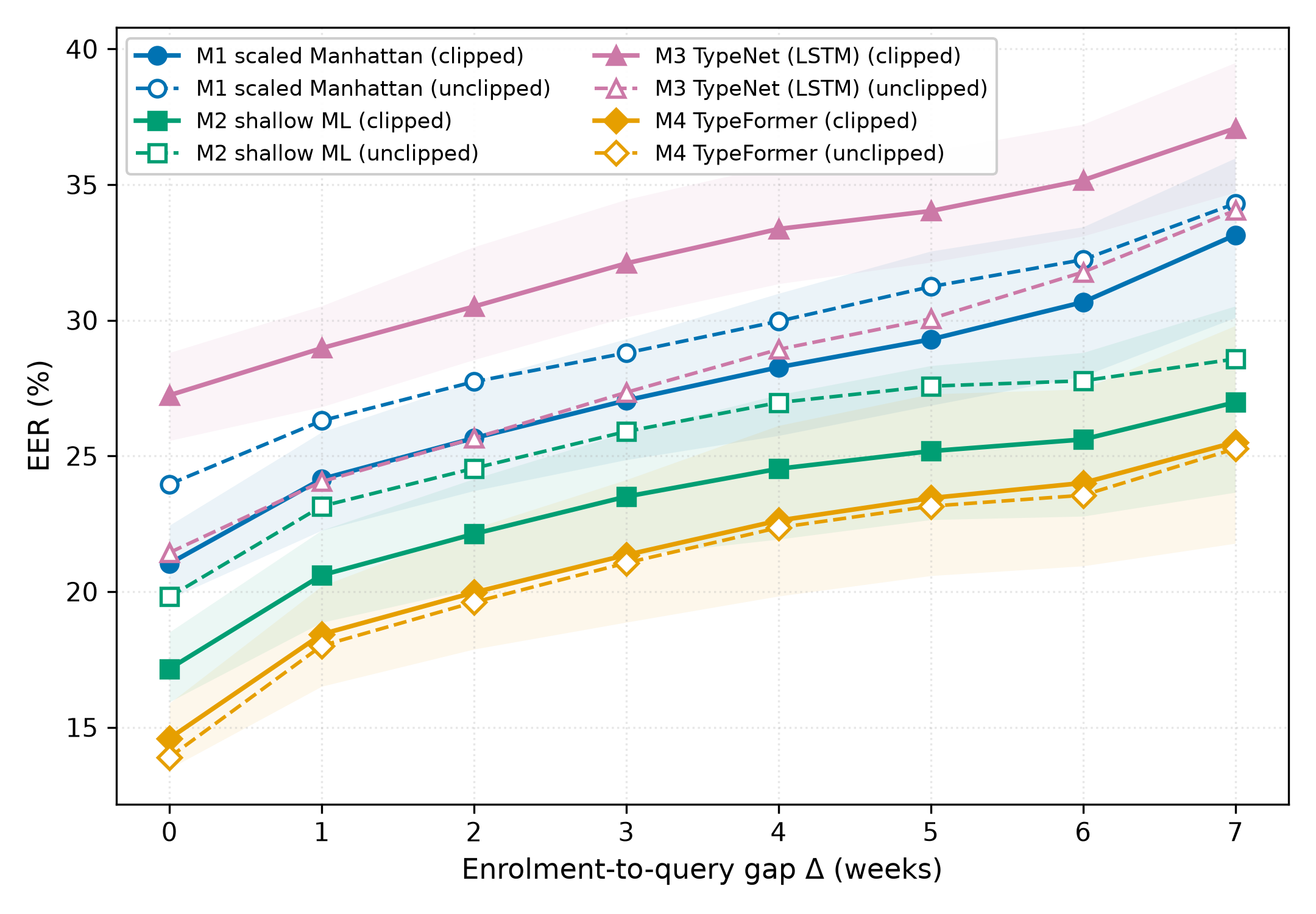}
\caption{EER versus enrolment-to-query gap $\Delta$ (weeks), by mechanism and outlier-clipping condition, at $k{=}5$ enrolment sessions. Shaded bands are bootstrap 95\% confidence intervals over test participants. Error increases monotonically with $\Delta$ for every mechanism.}
\label{fig:ageing}
\end{figure*}

Fig.~\ref{fig:ageing} plots the EER against the enrolment-to-query gap $\Delta$ for all four mechanisms, under both the outlier-clipped and unclipped preprocessing conditions, at the primary enrolment size $k{=}5$. Table~\ref{tab:eer} reports the corresponding numeric values at $\Delta{=}0$ and $\Delta{=}7$ with 95\% bootstrap CIs, alongside the genuine/impostor trial counts underlying every cell (identical across mechanisms at a given $\Delta$, since the trial design of Section~\ref{sec:trialdesign} depends only on which combinations (participant, password, week) exist, not on which mechanism scores them). Trial counts fall substantially as $\Delta$ grows, from 376{,}387 in total at $\Delta{=}0$ to 51{,}729 at $\Delta{=}7$, since fewer pairs of enrolment-week and query-week in the 8-week trial grid share a larger gap. This is also why the bootstrap confidence intervals widen visibly at $\Delta{=}7$. The full 8-point curve and per-fold breakdowns are provided in supplementary material.

\begin{table}[t]
\centering
\caption{EER (\%) at $\Delta{=}0$ and $\Delta{=}7$ ($k{=}5$, clipped, M3/M4 three-seed fusion, with 95\% bootstrap CI in brackets), together with the underlying genuine/impostor trial counts (identical across mechanisms).}
\label{tab:eer}
\begin{tabular}{lcc}
\toprule
Mechanism & EER at $\Delta{=}0$ & EER at $\Delta{=}7$ \\
\midrule
M1 (scaled Manhattan) & 21.0 [19.7, 22.4] & 33.1 [30.1, 36.0] \\
M2 (shallow ML)        & 17.2 [15.9, 18.5] & 27.0 [23.7, 30.5] \\
M3 (TypeNet-style LSTM) & 27.2 [25.6, 28.8] & 37.1 [34.7, 39.5] \\
M4 (TypeFormer-style)   & \textbf{14.6} [13.5, 15.9] & 25.5 [21.8, 29.8] \\
\midrule
\multicolumn{3}{l}{\textit{Trials (all mechanisms): genuine / impostor / total}} \\
$\Delta{=}0$ & \multicolumn{2}{l}{23{,}827 / 352{,}560 / 376{,}387} \\
$\Delta{=}7$ & \multicolumn{2}{l}{11{,}406 / 40{,}323 / 51{,}729} \\
\bottomrule
\end{tabular}
\end{table}

All mechanisms show a clear and monotonic increase in EER from $\Delta{=}0$ to $\Delta{=}7$, with relative increases ranging from 75\% for M4 to 58\% for M1. This pattern holds within each of the five test folds. $\Delta{=}7$ exceeds $\Delta{=}0$ in all five folds and in all four mechanisms, without exception. The finer adjacent-$\Delta$ step is not always an increase at the fold level, with small decreases in 2 of 35, 5 of 35, 0 of 35, and 5 of 35 adjacent-$\Delta$ steps for M1 through M4 respectively. This is consistent with sampling noise at the smaller per-fold trial counts, rather than genuine non-monotonicity. M4 achieves the lowest EER at every $\Delta$, followed by M2. M1 and M3 form a higher-error level, with M3 the worst throughout despite fusion. Absolute error rates exceed those of typical benchmarks for the same-session. Killourhy and Maxion report a 9.6\% EER for Manhattan~\cite{killourhy2009comparing}, compared to our 21.0\% for M1 at $\Delta{=}0$. Because M1 involves no training, the compute-budget argument discussed below cannot explain this gap, and we instead attribute it to differences in protocol, such as the cross-week design, the use of exhaustive impostors, and a different password corpus, without being able to isolate which factor dominates. For M3 and M4, a modest compute budget (Section~\ref{sec:limitations}) is an additional and separate factor.

\begin{figure*}[t]
\centering
\includegraphics[width=0.8\textwidth]{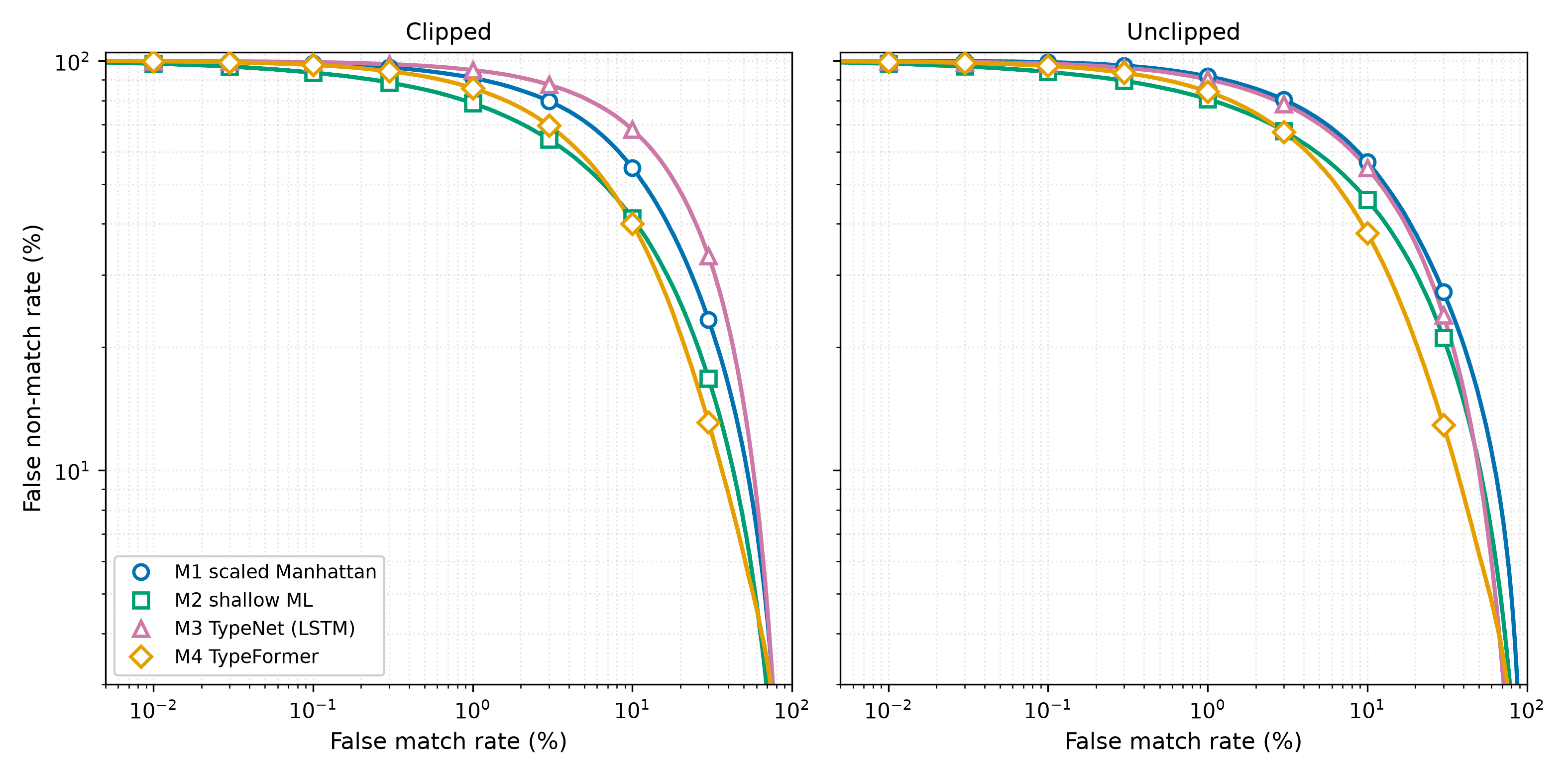}
\caption{DET curves at $k{=}5$, pooled across $\Delta$, for the outlier-clipped (left) and unclipped (right) conditions.}
\label{fig:det}
\end{figure*}

Fig.~\ref{fig:det} shows DET curves in $k{=}5$, pooled throughout $\Delta$, following ISO/IEC~19795-1~\cite{iso19795}. Table~\ref{tab:fnmr} reports FNMR at fixed-FMR operating points, and this is very high for all mechanisms, ranging from 69.6\% to 98.3\%. At the 0.1\%-FMR point in particular, all four mechanisms reject most genuine claims, which is a consequence of the same protocol-difficulty factors discussed above, plus the compute-budget factor for M3 and M4. This means that none of the four mechanisms would be usable at a low-FMR operating point without further improvement. The ordering here is mostly, though not perfectly, consistent with Table~\ref{tab:eer}. M3 remains the worst mechanism and M1 remains in the lower tier, but M2 and M4 swap places, with M2 achieving the lowest FNMR even though M4 achieves the lowest EER. This is expected rather than contradictory. EER is measured at each mechanism's own crossover point, whereas FNMR at a fixed FMR probes a different and stricter region of the DET curve, where the curves for two mechanisms can legitimately cross.

\begin{table}[t]
\centering
\caption{FNMR (\%) at fixed FMR operating points, $k{=}5$, clipped, pooled across $\Delta$.}
\label{tab:fnmr}
\begin{tabular}{lcc}
\toprule
Mechanism & FNMR @ FMR=1\% & FNMR @ FMR=0.1\% \\
\midrule
M1 (scaled Manhattan) & 91.5 & 98.6 \\
M2 (shallow ML)        & \textbf{79.7} & \textbf{93.9} \\
M3 (TypeNet-style LSTM) & 95.2 & 99.3 \\
M4 (TypeFormer-style)   & 86.4 & 97.8 \\
\bottomrule
\end{tabular}
\end{table}

\subsection{Ablations}
\label{sec:ablations}

Outlier clipping, using a 1.5$\times$IQR fence fit on training participants only, produces a small and consistent improvement for the classical and shallow-ML mechanisms, but has a more mixed, and in one case counter-intuitive, effect on the embedding models. M4 is essentially unaffected by clipping at $\Delta{=}0$, with 14.6\% clipped versus 13.9\% unclipped, while M3 is actually \emph{worse} when clipped than when unclipped at $\Delta{=}0$, at 27.2\% versus 21.4\%, although it is trained on the same raw sequence representation either way (Fig.~\ref{fig:ageing}, solid versus dashed lines). We do not have a confident explanation for M3's sensitivity to clipping, beyond noting that it is consistent with the broader finding, discussed in Section~\ref{sec:seedresults}, that M3 is the least stable of the four mechanisms. Full per-$\Delta$ clipping comparisons for all four mechanisms are given in the supplementary material.

We also swept the enrolment size $k \in \{1,3,5\}$, reporting the mean EER pooled in $\Delta$ under the clipped condition. M2 and M4 are close to saturated by $k{=}3$, with M2 moving from 22.7\% to 23.1\% to 23.2\% for $k=1,3,5$, and M4 moving from 23.0\% to 21.5\% to 21.2\%. M1 continues to improve noticeably with more enrolment sessions, from 33.3\% to 28.1\% to 27.4\%, while M3 is comparatively insensitive to $k$, moving only from 33.5\% to 32.7\% to 32.3\%. This is consistent with a model that has not learnt a strongly discriminative, sample-efficient representation, rather than one that is simply already saturated at $k{=}1$ in the way that M2 and M4 appear to be. The complete tables per-mechanism, per-$\Delta$ are given in the supplementary material.

\subsection{Mixed-Effects Model}
\label{sec:mixedeffects}

The full mixed-effects model (300{,}000 trials, stratified by mechanism and $\Delta$, drawn from the primary $k{=}5$, clipped configuration) converged cleanly. Fixed-effect estimates for the main-effects model are summarised in Table~\ref{tab:mixedmodel}. All reported effects are on the probability-of-decision-error scale.

\begin{table}[t]
\centering
\caption{Mixed-effects model: selected fixed-effect estimates (main-effects model, no interaction). $N=299{,}984$, with participant and password variance components of 0.003 and 0.001 respectively.}
\label{tab:mixedmodel}
\begin{tabular}{lccc}
\toprule
Term & Coef. & $z$ & $p$ \\
\midrule
$\Delta$ (per week, centred)         & $+0.017$ & 21.60 & $<0.001$ \\
Mechanism: M2 vs.\ M1                 & $-0.040$ & $-18.30$ & $<0.001$ \\
Mechanism: M3 vs.\ M1                 & $+0.052$ & $23.90$ & $<0.001$ \\
Mechanism: M4 vs.\ M1                 & $-0.065$ & $-30.15$ & $<0.001$ \\
Length (per character, centred)       & $-0.020$ & $-11.27$ & $<0.001$ \\
Hardware-change (week 5)              & $+0.005$ & $2.08$ & $0.037$ \\
\bottomrule
\end{tabular}
\end{table}

The ageing effect is confirmed with high confidence. Each additional week increases the probability of a decision error by 1.7 points, keeping all other covariates constant. All three alternative mechanisms differ significantly from M1, consistent with the order of EER in Table~\ref{tab:eer}. Longer passwords are associated with modestly lower error, and the week-5 hardware-change covariate, which reflects participants switching from university to personal computing equipment at the time of a national COVID-19 lockdown, is associated with a small but significant increase in error. The substitution category was also significant here, with $p<0.001$ for all four non-baseline categories, which was not the case in an earlier iteration of this analysis. We do not treat this particular result as stable, given the instability documented next and in Section~\ref{sec:ageingrates}.

The interaction between $\Delta$ and the mechanism, which tests whether the mechanisms age at a different rate than M1, is the central term of the model, and it is the term the three-seed fusion described in Section~\ref{sec:embedding} was introduced to make it trustworthy. In the primary fit, the joint test is not quite significant, with $\chi^2(3)=7.27$ and $p=0.064$. However, two of the three individual terms are significant. M2, with a coefficient of $-0.002$ and $p=0.029$, and M3, with a coefficient of $-0.002$ and $p=0.047$, both age significantly more slowly than M1, while M4, with a coefficient of $-0.000$ and $p=0.678$, does not differ from M1, despite having the best absolute accuracy. Resampling the same fitted set four different ways gives joint $p$-values of 0.064, 0.114, 0.015, and 0.021. This spread is smaller than the pre-fusion values of 0.157, 0.145, 0.020, and 0.021; however, resampling by itself is still insufficient to resolve the issue.

To test whether the interaction itself replicates under an entirely fresh run of the pipeline, using a different subject-level fold assignment (Section~\ref{sec:protocol}) and newly trained M1 through M4 for every fold, rather than simply a different resample of one fixed trial set, we repeated the full pipeline twice more end-to-end, each time with a different global random seed. Because the outcome is binary, the choice between a linear and a logistic specification discussed in Section~\ref{sec:statmodel} is a modelling decision we can test rather than a matter of presentation, so we also refit each replicate's interaction a second way, as a fixed-effects logistic regression with participant-clustered robust standard errors. 
Table~\ref{tab:replicates} reports both specifications side by side for all three runs. According to the linear-probability specification, the joint interaction test is significant in two of the three runs at $p<0.01$ and borderline in the third, with $\chi^2(3)=7.27, 17.92,$ and $16.22$, corresponding to $p=0.064, 0.0005,$ and $0.0010$. M2 and M3 both show a significantly shallower ageing slope than M1 in every run, while M4's confidence interval includes zero in every run. This pattern does not survive the logistic specification equally well. M3 remains negative and significant or near-significant in all three runs, with $p$ between 0.032 and 0.056. This is the one component of the original three-part claim that we consider robust to both replication and model specification. M2 loses significance according to the logistic specification in every run, with $p$ between 0.65 and 0.71 and an inconsistent sign. M4 shows the most striking reversal. Its logistic coefficient is consistently \emph{positive} and borderline to significant, with $p$ between 0.008 and 0.093, suggesting that it may in fact age \emph{faster} than M1 rather than at the same rate, the opposite conclusion to the null result of the linear model. A full crossed-random-effects binomial GLMM fitted via variational Bayes did not converge, but it stabilised on essentially the same point estimates as the cluster-robust fit, which reassures us that this pattern is not a numerical artefact of either method.

\begin{table*}[t]
\centering
\caption{$\Delta \times$ Mechanism interaction under the linear-probability and logistic (cluster-robust) specifications, across three independent end-to-end pipeline replicates. Cells show the coefficient, with the $p$-value in parentheses. In the linear specification, coefficients are expressed on the weekly probability-of-error scale per unit of $\Delta$, whereas in the logistic specification, coefficients are measured in log-odds. Consequently, only the direction and statistical significance of the coefficients, rather than their magnitudes, are directly comparable across the two model formulations. For the joint test in the linear model, the test statistics are $\chi^2(3)=7.27,\ 17.92,\ \text{and }16.22$, with corresponding $p$-values of $0.064,\ 0.0005,\ \text{and }0.0010$ for Runs 1, 2, and 3, respectively.}
\label{tab:replicates}
\begin{tabular}{lcccccc}
\toprule
 & \multicolumn{3}{c}{Linear probability model} & \multicolumn{3}{c}{Logistic (cluster-robust)} \\
Mechanism vs.\ M1 & Run 1 & Run 2 & Run 3 & Run 1 & Run 2 & Run 3 \\
\midrule
M2 & $-0.002$ (.029) & $-0.003$ (.010) & $-0.002$ (.030) & $-0.004$ (.652) & $-0.004$ (.650) & $+0.003$ (.713) \\
M3 & $-0.002$ (.047) & $-0.004$ ($<$.001) & $-0.004$ (.001) & $-0.019$ (.032) & $-0.021$ (.016) & $-0.015$ (.056) \\
M4 & $-0.000$ (.678) & $-0.001$ (.485) & $-0.000$ (.954) & $+0.016$ (.093) & $+0.016$ (.090) & $+0.024$ (\textbf{.008}) \\
\bottomrule
\end{tabular}
\end{table*}

Statistical significance is not the same as practical importance. Even where they are significant, the linear-model coefficients are small, ranging from $-0.002$ to $-0.004$ per week. Across the seven-week span this amounts to the error probability for M2 and M3 rising 1.4 to 2.8 percentage points \emph{less} than M1's, which agrees with the raw differences in Table~\ref{tab:eer}, where M1 rises by 12.1 points against 9.8 for M2 and 9.9 for M3. The gap to M1 therefore narrows, but the accuracy ranking is preserved at every value of $\Delta$. The M2 estimate should still be read with caution given the logistic results above. We do not read the divergence between the linear and logistic estimates as a sign that the data are unreliable. Sign and significance need not agree across specifications when baseline rates differ substantially, as they do here between M4's $14.6\%$ and M3's $27.2\%$, and this is a well-documented form of scale sensitivity rather than a coding error. One further caveat is worth stating plainly. We ran replicates 2 and 3 because run 1 was borderline, not because the number of replicates was fixed in advance, which is a researcher degree of freedom that a pre-registered plan would have avoided. The individual M2 and M3 coefficients were already nominally significant before that decision, however, and their signs are stable across all three runs. Section~\ref{sec:ageingrates} revises the headline claim accordingly.

\subsection{Training-Seed Robustness}
\label{sec:seedresults}

Table~\ref{tab:seedrobustness} reports the result of the training-seed robustness check described in Section~\ref{sec:seedcheck}. Each fold's M3 and M4 were retrained three times on identical data with different random seeds.

\begin{table}[t]
\centering
\caption{Decomposition of fold-to-fold EER variance into training-seed noise (same data, different training run) versus total across-fold spread.}
\label{tab:seedrobustness}
\begin{tabular}{lccc}
\toprule
Mechanism & Within-fold seed SD & Across-fold SD & Noise share \\
\midrule
M3 (LSTM)        & 0.0181 & 0.0313 & $\sim$58\% \\
M4 (Transformer) & 0.0062 & 0.0325 & $\sim$19\% \\
\bottomrule
\end{tabular}
\end{table}

Fifty-eight percent of M3's fold-to-fold EER variance is attributable to training-run randomness alone, with the data held fixed. M4 is much more stable, at only 19\%. This directly explains why the interaction estimate discussed above is sensitive to which specific training run of M3 entered the analysis, and why we do not report the interaction as confirmed on the strength of a single training run per fold.

This pattern is visible not only as a summary statistic. For each of the five folds, the three independent, unfused M3 training runs are visibly more spread out than the three M4 runs, and in several folds M3's per-fold spread is comparable to, or even larger than, the difference between folds (the full per-fold breakdown is given in the supplementary material).

\section{Discussion}
\label{sec:discussion}

The following three findings are robust across all versions of this analysis, including earlier corrected iterations, and hold under both linear and logistic mixed-model specifications described in Section~\ref{sec:mixedeffects}.

\begin{itemize}
\item The verification error increases monotonically and significantly with $\Delta$ for each mechanism, confirming that the ageing of the template is a real and measurable effect over eight weeks.
\item The four mechanisms differ significantly in the accuracy of the baseline, and M4 and M2 materially outperform M1 and M3.
\item Password length and the hardware-change covariate both have small but statistically detectable effects, which justifies including them rather than holding them implicitly constant.
\end{itemize}

The fourth finding is more qualified. M3 ages significantly more slowly than M1 in every check we performed (Section~\ref{sec:ageingrates}), but whether M2 shares this property and whether M4 truly ages at the same rate as M1 rather than faster are both specification-dependent rather than settled.

\subsection{Mechanism-Dependent Ageing Rates}
\label{sec:ageingrates}

The paper has a conceptual contribution beyond the ageing curve itself. Absolute verification accuracy and ageing robustness are separate properties of a matching mechanism, and a designer who selects a mechanism on accuracy alone may be optimising the wrong quantity for a system that will be used for more than a few weeks without re-enrolment. Whether the ageing \emph{rate} depends on the mechanism, the question raised in Section~\ref{sec:trialdesign}, needed more scrutiny than any single check could provide. The honest answer turns out to be narrower than our earlier drafts claimed. Of the three mechanism-specific effects that we originally identified, only one is robust across every check we ran, and the other two are not.

Section~\ref{sec:mixedeffects} and Table~\ref{tab:replicates} established this pattern. M3 ages significantly more slowly than M1 under both the linear-probability and the logistic specifications, and in all three replicates, which is the one component of our original claim that we consider well supported. The apparent advantage of M2 and the apparent parity of M4, by contrast, are artefacts of the linear specification that do not survive a logistic one. M2 loses significance and M4 reverses sign, suggesting that it may, in fact, age faster. This does not mean that the underlying data are unreliable. This kind of scale-dependence, sometimes called non-collapsibility, is expected when the groups being compared have substantially different baseline rates, as they do here. M3's result is the stronger claim precisely because it survives both the fold-replication check and this change in the functional form of the model.

This leaves a narrower decoupling. M3, the least accurate mechanism, is the one whose slower ageing is robust to the choice of outcome model, while M4, the most accurate, shows tentative signs of ageing faster instead. We lack a confirmed mechanistic explanation for this, but we can offer one testable hypothesis, which remains explicitly unverified. Under this floor-effect account, if much of M3's error, even at $\Delta{=}0$, stems from limited representational quality (Section~\ref{sec:limitations}) rather than from genuine drift, then M3 has less remaining ``headroom'' left to grow as $\Delta$ increases. The much lower starting error of M4, by contrast, leaves more room for real degradation to appear as $\Delta$ grows, which would be consistent with, although it does not prove, the suggestion of the logistic model that M4 ages faster. Under this account, M3's shallow slope would reflect poor optimisation rather than a general property of recurrent architectures, and it would be expected to steepen with better training. This is a specific target for follow-up work (Section~\ref{sec:limitations}), rather than a conclusion that our present data can support on its own.

We consider this process, and not only the final numbers, to be the paper's real contribution. A biometrics literature that reports mechanism comparisons from a single training run, without a seed-variance decomposition, an independent replication, and an outcome-model sensitivity check, risks reporting exactly the kind of artefact-prone result that two of our three original sub-claims turned out to be. 

\subsection{Limitations}
\label{sec:limitations}

\textbf{Training regime and data scale.} M3 and M4 were trained on a single consumer-grade GPU, which is a modest budget compared with the original TypeNet and TypeFormer training regimes. A distinct and more fundamental limitation is the population size. TypeNet and TypeFormer are designed to be trained on thousands of subjects, and our 85-participant pool is one to two orders of magnitude smaller than that. More GPU-hours would not fix this on their own. We cannot rule out that M3's poor absolute accuracy (Table~\ref{tab:eer}) reflects too small a population from which to learn a discriminative general representation, independent of training compute. Together with the genuine difficulty of our cross-week forced-impostor protocol, which, to our knowledge, no other fixed-text study has attempted, these three factors plausibly explain the gap to the sub-10\% EER figures reported for classical matchers elsewhere~\cite{killourhy2009comparing}, although we cannot separate their individual contributions.

\textbf{Mechanism comparability.} M2 trains 40 separate per-password classifiers, each solving an easier, password-specific problem, whereas M3 and M4 share a single representation across all passwords. This may benefit M2 relative to a deployment using a single generic model, and we did not test a population-level version of M2 to isolate this effect. We also evaluate only one recurrent and one attention-based architecture, rather than the broader family of self-supervised, contrastive, Siamese, and temporal-convolutional approaches (Section~\ref{sec:mechanisms}). We narrowed the scope this way so that the three-seed fusion and three-replicate validation described in Section~\ref{sec:ageingrates}, which we consider essential rather than optional, remained tractable. Whether M3's advantage in ageing rate reflects recurrent architectures in general, or is specific to this particular accuracy gap, is accordingly still an open question. No hyperparameter search was performed for M3 or M4.

\textbf{Modelling choices.} The choice between a linear and a logistic specification is no longer hypothetical, as discussed in Section~\ref{sec:statmodel}. Section~\ref{sec:mixedeffects} shows that it materially changes two of the three interaction conclusions. Our logistic check used cluster-robust fixed effects rather than an exact random-effects analogue, since a full binomial GLMM did not converge at this scale. A converged GLMM, an \texttt{lme4} fit, or a larger dataset might still resolve M2 and M4 differently. We also have not checked whether secondary covariates, such as substitution category, which flipped significance between analysis iterations, are as stable across replicates as the central interaction is, and we recommend that check before treating any secondary effect as more than suggestive.

\textbf{Population and deployment scope.} The 85-participant pool comes from a single institution, and we have no demographic breakdown by age, handedness, proficiency, or keyboard layout with which to test generalisation. That said, the qualitative ageing effect, which is consistent across all four independently-implemented mechanisms, is the focus of this investigation, rather than the minimisation of absolute error rates. The 40-password corpus is shared across participants rather than user-chosen, which is standard for controlled studies but is not representative of self-selected deployment passwords. Our results use a threshold that is recalibrated for each fold and each $\Delta$, rather than a single fixed threshold held constant across a deployment's lifetime, so our curves characterise how separable genuine and impostor samples are at each gap, rather than forecasting the error rate of a fixed-threshold system over time. 

\section{Conclusion}
\label{sec:conclusion}

Using a unified genuine and impostor trial design that jointly varies matching mechanism and enrolment-to-query gap, we show that fixed-text keystroke verification error grows significantly and substantially over an eight-week span, for a classical distance-threshold matcher, a shallow-ML classifier, and two learnt sequence-embedding models alike. This ageing effect, rather than the mechanism comparison that follows, is the paper's most robust empirical claim. The four mechanisms differ significantly in baseline accuracy, with a TypeFormer-style Transformer embedding achieving the lowest error throughout. Whether they also age at different rates is a harder question, and our own answer to it changed as we subjected it to more scrutiny. Three seemingly confirmed mechanism-specific effects, having survived seed fusion and three independent pipeline replications, dropped to just one once tested against a second, equally defensible outcome-model specification. Only the recurrent embedding model's slower ageing rate survives every check. The shallow-ML classifier's apparent advantage and the Transformer model's apparent parity are both artefacts of the linear-probability specification, and under a logistic one the Transformer instead shows tentative signs of ageing faster. We view this progressive narrowing, rather than the final mechanism ranking, as the paper's central methodological contribution. We recommend seed fusion, independent replication, and outcome-model sensitivity checking together, rather than individually, as standard practice whenever a learnt biometric matcher's comparative ranking or rate of change is reported.

\section{Appendix}

\subsection{Data Availability}
All the data and code collected and written and used in this manuscript is available at: \url{https://github.com/sparkins01/keystroke}

\subsection{Password Corpus}

The following list provides the lists of password phrases used in this research.

\begin{enumerate}
 \item action
\item return
\item bacteria
\item football
\item calculated
\item automotive
\item professional
\item technologies
\item Filter
\item docTor
\item coMputer
\item clickiNg
\item conDitions
\item Conference
\item disappointeD
\item inflaMmation
\item brok3n
\item cr1sis
\item deliv3ry
\item ann0ying
\item underst0od
\item addressin9
\item headqu4rters
\item pr3scription
\item fr!end
\item gard£n
\item d$|$ameter
\item rec\$ives
\item univers!ty
\item de\textbackslash{}ivering
\item bre\$thtaking
\item embarrassin?
\item F@st3r
\item pOl!c3
\item sc!enC3
\item he4Ven$|$y
\item $|$ndig3nOus
\item in5ul@tIon
\item aSynchr0\#ous
\item cat@s7Rophic
\end{enumerate}

\subsection{Full Ablation Results}

This appendix gives the complete per-$\Delta$ breakdowns summarised in the main text's Ablations and Evaluation Metrics sections, plus a qualitative note on a negative-mining ablation that was resolved during development rather than reported as a headline result.

\subsubsection{Trial Counts, All Eight Gaps}

Table~\ref{tab:trialcounts} gives the genuine and impostor trial counts underlying the main text's EER table and ageing-curve figure, for all eight values of $\Delta$ (identical across mechanisms at a given $\Delta$, since the trial design depends only on which participant/password/week combinations exist).

\begin{table}[t]
\centering
\caption{Genuine and impostor trial counts by $\Delta$ ($k{=}5$, clipped; identical across mechanisms).}
\label{tab:trialcounts}
\begin{tabular}{lccc}
\toprule
$\Delta$ & Genuine trials & Impostor trials & Total \\
\midrule
0 & 23{,}827  & 352{,}560 & 376{,}387 \\
1 & 79{,}994  & 293{,}714 & 373{,}708 \\
2 & 68{,}527  & 249{,}512 & 318{,}039 \\
3 & 56{,}387  & 203{,}999 & 260{,}386 \\
4 & 45{,}332  & 163{,}875 & 209{,}207 \\
5 & 34{,}393  & 123{,}530 & 157{,}923 \\
6 & 22{,}474  & 80{,}251  & 102{,}725 \\
7 & 11{,}406  & 40{,}323  & 51{,}729  \\
\bottomrule
\end{tabular}
\end{table}

\subsubsection{Outlier Clipping, All Eight Gaps}

Table~\ref{tab:appclip} extends the main text's EER table and ageing-curve figure to all eight values of $\Delta$ and both clipping conditions. The pattern noted in the main text, where clipping helps M1 and M2 modestly and consistently but \emph{hurts} M3 and is roughly neutral for M4, holds at every $\Delta$ and not only at the two endpoints reported there.

\begin{table*}[t]
\centering
\caption{Full EER (\%) by mechanism, clipping condition, and temporal gap $\Delta$, $k{=}5$.}
\label{tab:appclip}
\begin{tabular}{llcccccccc}
\toprule
Mechanism & Condition & $\Delta{=}0$ & $\Delta{=}1$ & $\Delta{=}2$ & $\Delta{=}3$ & $\Delta{=}4$ & $\Delta{=}5$ & $\Delta{=}6$ & $\Delta{=}7$ \\
\midrule
\multirow{2}{*}{M1 (scaled Manhattan)} & Clipped   & 21.0 & 24.2 & 25.7 & 27.0 & 28.3 & 29.3 & 30.7 & 33.1 \\
                                       & Unclipped & 24.0 & 26.3 & 27.7 & 28.8 & 30.0 & 31.2 & 32.2 & 34.3 \\
\multirow{2}{*}{M2 (shallow ML)}       & Clipped   & 17.2 & 20.6 & 22.1 & 23.5 & 24.5 & 25.2 & 25.6 & 27.0 \\
                                       & Unclipped & 19.8 & 23.1 & 24.5 & 25.9 & 27.0 & 27.6 & 27.8 & 28.6 \\
\multirow{2}{*}{M3 (TypeNet-style LSTM)} & Clipped & 27.2 & 29.0 & 30.5 & 32.1 & 33.4 & 34.0 & 35.2 & 37.1 \\
                                       & Unclipped & 21.4 & 24.1 & 25.7 & 27.3 & 28.9 & 30.1 & 31.8 & 34.1 \\
\multirow{2}{*}{M4 (TypeFormer-style)} & Clipped   & 14.6 & 18.4 & 20.0 & 21.4 & 22.6 & 23.4 & 24.0 & 25.5 \\
                                       & Unclipped & 13.9 & 18.0 & 19.6 & 21.1 & 22.4 & 23.1 & 23.6 & 25.3 \\
\bottomrule
\end{tabular}
\end{table*}

\subsubsection{Enrolment Burden, All Eight Gaps}

Table~\ref{tab:appksweep} extends the main text's $k$-sweep summary to all eight values of $\Delta$, clipped condition. M1's improvement from more enrolment sessions and M3's comparative insensitivity to $k$ are visible at every $\Delta$, not only in the pooled average.

\begin{table*}[t]
\centering
\caption{Full EER (\%) by mechanism, enrolment size $k$, and temporal gap $\Delta$, clipped condition.}
\label{tab:appksweep}
\begin{tabular}{llcccccccc}
\toprule
Mechanism & $k$ & $\Delta{=}0$ & $\Delta{=}1$ & $\Delta{=}2$ & $\Delta{=}3$ & $\Delta{=}4$ & $\Delta{=}5$ & $\Delta{=}6$ & $\Delta{=}7$ \\
\midrule
\multirow{3}{*}{M1 (scaled Manhattan)} & 1 & 28.9 & 31.4 & 32.3 & 33.2 & 34.0 & 34.6 & 35.0 & 36.5 \\
                                       & 3 & 21.9 & 25.6 & 26.5 & 27.9 & 29.2 & 30.0 & 30.6 & 33.1 \\
                                       & 5 & 21.0 & 24.2 & 25.7 & 27.0 & 28.3 & 29.3 & 30.7 & 33.1 \\
\multirow{3}{*}{M2 (shallow ML)}       & 1 & 15.8 & 20.5 & 22.0 & 23.3 & 24.2 & 24.8 & 24.7 & 26.5 \\
                                       & 3 & 16.1 & 20.7 & 22.1 & 23.5 & 24.8 & 25.2 & 25.6 & 26.9 \\
                                       & 5 & 17.2 & 20.6 & 22.1 & 23.5 & 24.5 & 25.2 & 25.6 & 27.0 \\
\multirow{3}{*}{M3 (TypeNet-style LSTM)} & 1 & 29.9 & 31.6 & 32.5 & 33.5 & 34.3 & 34.9 & 34.9 & 36.6 \\
                                       & 3 & 27.6 & 29.7 & 31.0 & 32.5 & 33.8 & 34.6 & 35.2 & 37.1 \\
                                       & 5 & 27.2 & 29.0 & 30.5 & 32.1 & 33.4 & 34.0 & 35.2 & 37.1 \\
\multirow{3}{*}{M4 (TypeFormer-style)} & 1 & 16.2 & 20.8 & 21.8 & 23.3 & 24.6 & 25.1 & 25.5 & 26.7 \\
                                       & 3 & 14.1 & 19.0 & 20.2 & 21.8 & 23.3 & 24.1 & 24.3 & 25.5 \\
                                       & 5 & 14.6 & 18.4 & 20.0 & 21.4 & 22.6 & 23.4 & 24.0 & 25.5 \\
\bottomrule
\end{tabular}
\end{table*}

\subsubsection{Participant-Balanced EER Sensitivity Check}

Table~\ref{tab:particbalanced} reports the full per-mechanism, per-$\Delta$ comparison referenced in the main text's Evaluation Metrics section: pooled-trial EER (as reported throughout the main text) versus a per-participant-balanced EER, computed as one EER per genuine-claimant test participant (that participant's own genuine trials against the full pooled impostor set for that mechanism-$\Delta$ cell), averaged unweighted across participants. The two estimators agree closely at every $\Delta$ for every mechanism, with the largest discrepancy 1.7 percentage points (M2, $\Delta{=}4$); the mechanism ordering (M4 best, then M2, then M1, then M3) is identical under both estimators at every $\Delta$. We conclude that the pooled-trial-weighting convention adopted for consistency with ISO/IEC~19795-1 does not materially affect any comparative claim made in this paper.

\begin{table*}[t]
\centering
\caption{EER (\%) at $k{=}5$, clipped: pooled-trial vs.\ per-participant-balanced estimator, all eight values of $\Delta$.}
\label{tab:particbalanced}
\begin{tabular}{llcccccccc}
\toprule
Mechanism & Estimator & $\Delta{=}0$ & $\Delta{=}1$ & $\Delta{=}2$ & $\Delta{=}3$ & $\Delta{=}4$ & $\Delta{=}5$ & $\Delta{=}6$ & $\Delta{=}7$ \\
\midrule
\multirow{2}{*}{M1 (scaled Manhattan)} & Pooled            & 21.0 & 24.2 & 25.7 & 27.0 & 28.3 & 29.3 & 30.7 & 33.1 \\
                                       & Particip.-balanced & 20.9 & 24.2 & 25.6 & 26.7 & 28.2 & 29.4 & 30.8 & 32.2 \\
\multirow{2}{*}{M2 (shallow ML)}       & Pooled            & 17.2 & 20.6 & 22.1 & 23.5 & 24.5 & 25.2 & 25.6 & 27.0 \\
                                       & Particip.-balanced & 16.2 & 19.2 & 20.7 & 21.9 & 22.9 & 23.8 & 24.2 & 26.0 \\
\multirow{2}{*}{M3 (TypeNet-style LSTM)} & Pooled          & 27.2 & 29.0 & 30.5 & 32.1 & 33.4 & 34.0 & 35.2 & 37.1 \\
                                       & Particip.-balanced & 27.0 & 28.6 & 30.2 & 31.6 & 32.8 & 33.7 & 34.6 & 37.7 \\
\multirow{2}{*}{M4 (TypeFormer-style)} & Pooled            & 14.6 & 18.4 & 20.0 & 21.4 & 22.6 & 23.4 & 24.0 & 25.5 \\
                                       & Particip.-balanced & 13.8 & 17.7 & 19.3 & 20.6 & 22.1 & 23.2 & 23.8 & 26.7 \\
\bottomrule
\end{tabular}
\end{table*}

\subsubsection{Training-Seed Robustness, Full Per-Fold Breakdown}

The main text's Training-Seed Robustness section summarises the training-seed variance decomposition as within-fold and across-fold standard deviations. Table~\ref{tab:seedfold} gives the underlying raw values: pooled EER for each of three independently-trained (unfused) seeds of M3 and M4, in each of the five test folds. The spread of M3 within the fold across the seeds is visibly wider than that of M4 in every fold, and in several folds (e.g.\ fold~0, fold~2) is comparable to or larger than the spread between folds.

\begin{table}[t]
\centering
\caption{Pooled EER (\%) for three independently-trained (unfused) seeds of M3 and M4, by test fold.}
\label{tab:seedfold}
\begin{tabular}{lcccc}
\toprule
Fold & Mechanism & Seed 0 & Seed 1 & Seed 2 \\
\midrule
\multirow{2}{*}{0} & M3 & 29.7 & 35.2 & 29.6 \\
                    & M4 & 20.6 & 20.8 & 21.5 \\
\multirow{2}{*}{1} & M3 & 31.0 & 34.5 & 33.7 \\
                    & M4 & 18.9 & 17.7 & 17.2 \\
\multirow{2}{*}{2} & M3 & 35.5 & 32.5 & 32.6 \\
                    & M4 & 18.6 & 19.4 & 21.0 \\
\multirow{2}{*}{3} & M3 & 24.8 & 27.6 & 25.4 \\
                    & M4 & 19.5 & 19.8 & 19.9 \\
\multirow{2}{*}{4} & M3 & 33.9 & 32.2 & 32.7 \\
                    & M4 & 26.6 & 26.1 & 26.7 \\
\bottomrule
\end{tabular}
\end{table}

\subsubsection{Negative-Mining Strategy for the Triplet Loss (M3/M4 Training)}

The main text states that the embedding models are trained with semi-hard-negative mining and notes in passing that pure hardest-negative mining was tried first and rejected. We record the full observation here, since it may be useful to others implementing similar triplet-loss training on small keystroke-dynamics datasets. With hardest-negative mining (always selecting, from a candidate pool, the impostor sample closest to the anchor), training loss collapsed to exactly the margin value within a single epoch and did not move thereafter, for both M3 and M4. Inspecting the resulting embeddings confirmed the classic embedding-collapse failure mode: all outputs converged to (approximately) the same point in the embedding space, which trivially satisfies a triplet margin loss when every pairwise distance is near zero. This is a known risk of hardest-negative mining early in training, when the embedding space has no learnt structure yet to make the mining meaningful, and is why the original FaceNet formulation of semi-hard mining (selecting the closest negative that is still farther from the anchor than the positive, falling back to the hardest available negative only when no candidate qualifies) exists as a standard alternative. Switching to semi-hard mining immediately resolved the collapse, and all results reported in the main text use semi-hard mining throughout; no quantitative comparison table is given because the hardest-mining configuration never produced a trained model worth evaluating.


\bibliographystyle{IEEEtran}
\bibliography{references}

\end{document}